\documentclass[12pt]{article}

\usepackage{makeidx}
\usepackage{graphicx} 
\usepackage{subeqnar} 
\usepackage{multicol} 
\usepackage{cropmark} 

\begin{document}
\title{A characteristic time scale of tick quotes on foreign currency markets}
\author{Aki-Hiro Sato \\
Department of Applied Mathematics and Physics, \\
Graduate School of Informatics, \\
Kyoto University, Kyoto 606-8501, Japan
}

\maketitle              

\begin{abstract}
This study investigates that a characteristic time scale on an exchange 
rate market (USD/JPY) is examined for the period of 1998 to 2000. 
Calculating power spectrum densities for the number
of tick quotes per minute and averaging them over the year yield that 
the mean power spectrum density has a peak at high frequencies. 
Consequently it means that there exist the characteristic scales which
dealers act in the market. A simple agent model to explain this
phenomenon is proposed. This phenomena may be a result of stochastic
resonance with exogenous periodic information and physiological
fluctuations of the agents. This may be attributed to the traders'
behavior on the market. The potential application is both quantitative
characterization and classification of foreign currency markets.

\end{abstract}

\section{Introduction}
Empirical analysis of high-frequency financial data have been
attracting significant interest among physicists as well as economists
during a decade (Mantegna and Stanley 2000, and Dacorogna 2001). Many 
features of financial markets have been clarified by many successive studies.

Actually it is well-known that the markets have a characteristic time
scale in long period (daily, weekly, and monthly). However recent
studies (Takayasu 2003, Ohnishi 2004 and Mizuno 2004) on time-series
analyses in financial markets show that the market has a
characteristic time scale in short period and propose the reason why
traders are mainly using strategies with weighted feedbacks of past
prices. Furthermore using the self-modulation process Takayasu {\it et
  al.} have found that the characteristic time scale is about 2 minutes
in the JPY/USD market (Takayasu 2003) (abbreviated as MT). 

On the other hand Baninec and Krawiech and Ho\l st proposed a possibility 
that stochastic resonance occurs in markets (Babinec 2002 and Krawiech
2003) through an Ising-like agent model. They suggest that a
periodicity in the market results from exogenous periodical
information (abbreviated as BKH). 

In order to clarify the mechanism of this characteristic time I think
that we should examine it on a different standpoint from MT and BKH.
Both studies focus on prices or price returns. However, in this
article, we focus on the number of tick quotes in foreign currency
rates (USD/JPY) and investigate the statistical properties of 
them by utilizing the power spectrum technique. As the results of
examining the number of tick quotes in USD/JPY market it is found that
the power spectrum density (PSD) has some peaks at about 2 minutes
(the peak frequency depends on the currency markets).

In order to explain this phenomena a simple agent model based on
double-threshold noisy device (Sato 2004) is proposed. From a
result of numerical simulations of the model it is found that the high
periodicity of the number of tick quotes may happen. This result leads
to a hypothesis that this periodicity is caused by common exogenous
periodical information.

The purposes of this study are as follows: (1) to examine the number
of high-frequency quotes lead us to deeply understand microscopic
market activities. (2) this may provide useful information for market
players to consider their trading strategy.

\section{Data Analysis}
\begin{figure}[t]
\centering
\includegraphics[scale=0.3,angle=270]{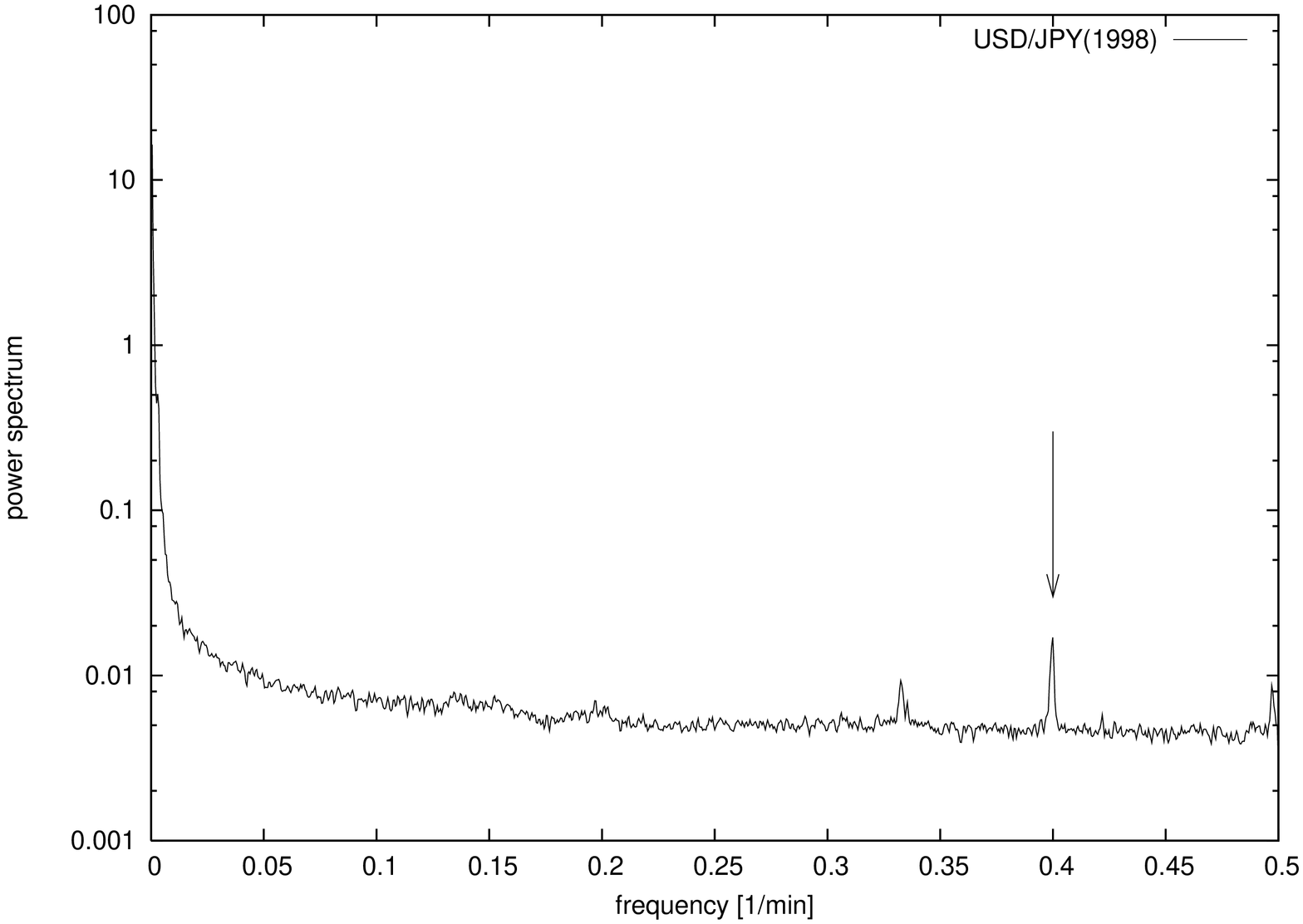}
\includegraphics[scale=0.3,angle=270]{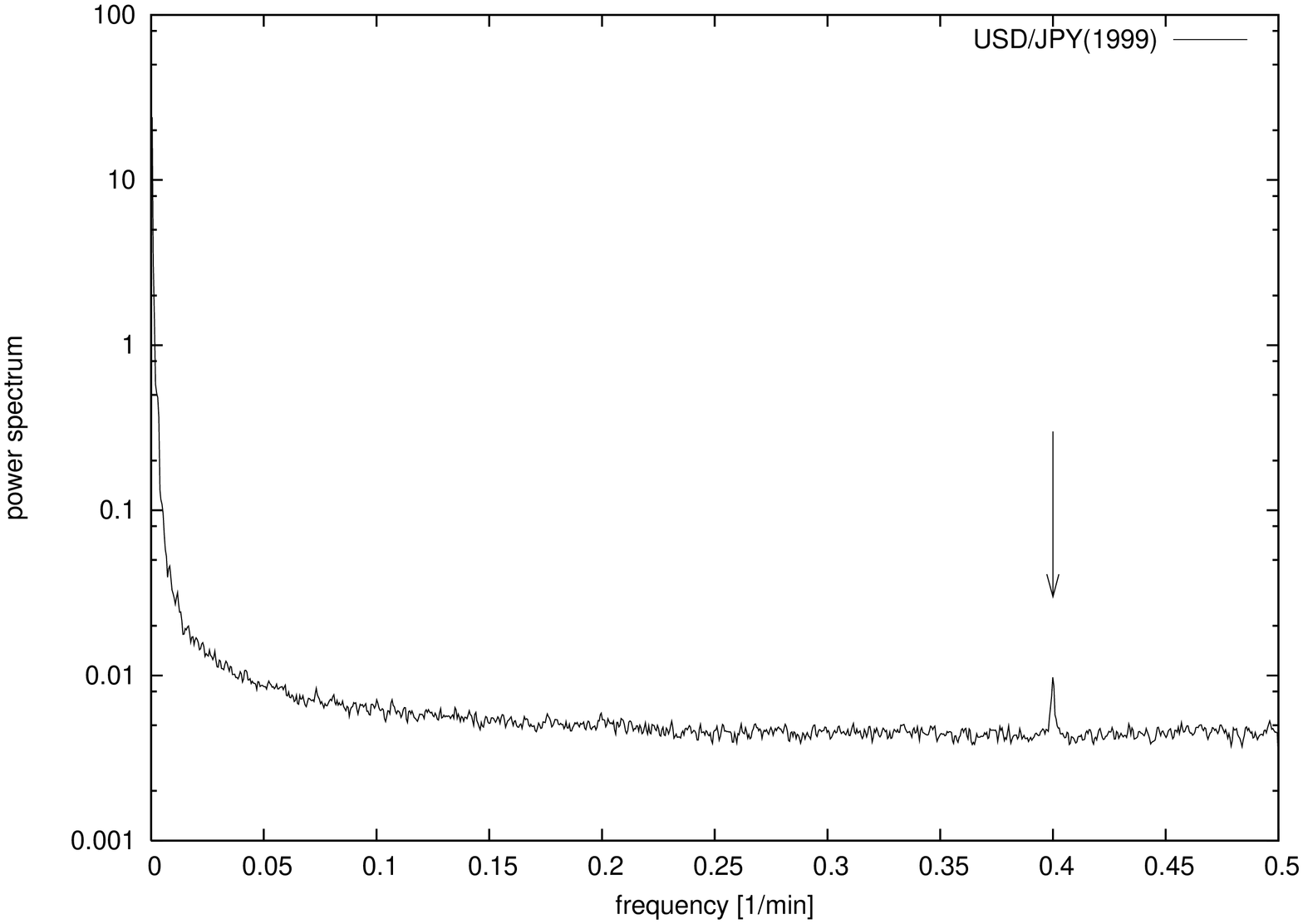}
\includegraphics[scale=0.3,angle=270]{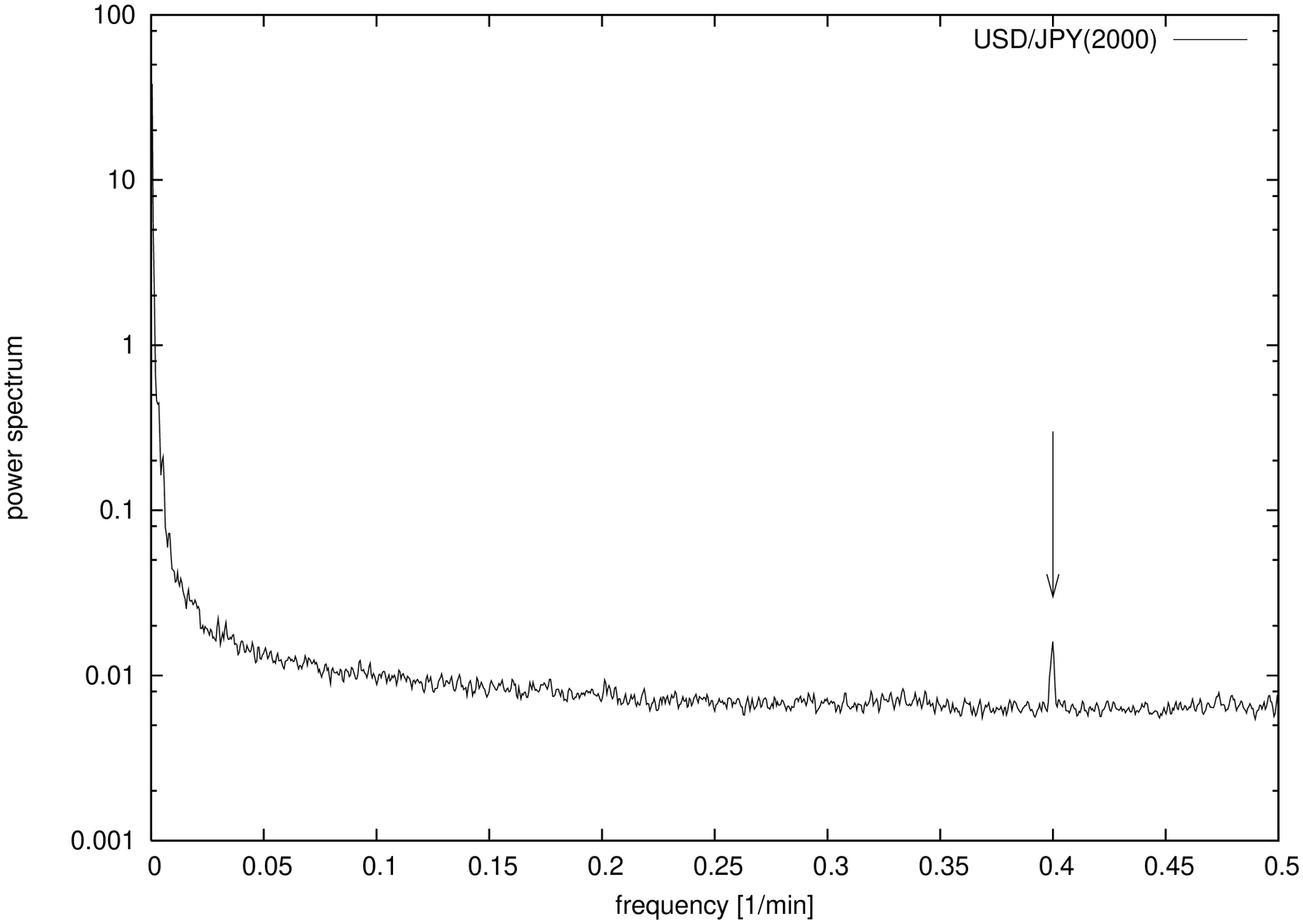}
\caption{Semi-log plots of the averaged power spectrum density 
of the number of tick quotes (USD/JPY) over the year for a period of 
1998 (top), 1999 (middle), and 2000 (bottom). They all have a peak 
at 0.4 (1/min), i.e., 2.5 minutes.}
\label{fig:powerspectrum}
\end{figure}

The number of ask quotes per minute in USD/JPY is counted for a period
of 1998 to 2000. Utilizing the data we calculated three PSDs for 2,048
points in weekday and average them over the year. The averaged power
spectrums on the semi-log scale are shown in
Fig. \ref{fig:powerspectrum}. They all have a peak at 0.4 (1/min),
namely 2.5 minutes. We consider that these peaks exhibit
characteristic time scales of dealers' activities, i.e., the dealers act
having the periodicity of 2.5 minutes.

\section{Dealer model} 
We introduce a simple agent model based on double-threshold noisy
devices in order to understand the characteristic time scales found in
the averaged power spectrum density. 
This model contains $N$ dealers and the $i$th dealer has
double-threshold ($\theta_i^{(1)} > \theta_i^{(2)}$) to decide buy(1),
sell(-1) and wait(0), and noise source $\xi_i(t)$ to model an
uncertainty in their mind. We assume that the $i$th dealer must choose
a decision (output) into the three ones $y_i(t)=\{1,0,-1\}$ based on
information (input) $x_i(t)$ with an uncertainty $\xi(t)$ in his/her mind:

\begin{equation}
y_i(t) =
\left\{
\begin{array}{ll}
1 & (x_i(t) + \xi_i(t) > \theta_i^{(1)}) \\
0 & (\theta_i^{(2)} \leq x_i(t) + \xi_i(t) \leq \theta_i^{(1)}) \\
-1 & (x_i(t) + \xi_i(t) < \theta_i^{(2)})
\end{array}
\right..
\end{equation}
Here we assume that $\xi_i(t)$ is identically independent Gaussian
distribution,

\begin{equation}
G(\xi) =
\frac{1}{\sqrt{2\pi}\sigma_i}\exp\Bigr(-\frac{\xi^2}{2\sigma_i^2}\Bigl),
\end{equation}
where $\sigma_i$ are standard deviations of the $i$th dealer.

It is assumed that the input of each dealer is exogenous periodic
information $s(t)=A \sin(2 \pi f t)$, where $A$ represents an
amplitude, and $f$ a frequency. For $s(t)>0$ the dealers feel it good 
news and tend to decide a buy, while for $s(t)<0$ they do it bad news
and to decide a sell. 

Furthermore the number of tick quotes per unit time $X(t)$ is defined as  
\begin{equation}
X(t) = \frac{1}{N}\sum_{i=1}^{N}|y_i(t)|.
\end{equation}

For simplicity assume $\theta_i^{(1)} = \theta$ and $\theta_i^{(2)} =
-\theta$. Obviously the activity $X(t)$ is always zero if $\sigma=0$
and $A<\theta$, so that, there is no uncertainty of the dealers in
their mind and the exogenous information is weaker than the threshold
for the dealers to decide their action. However if there is
uncertainty $\sigma>0$ then the activity $X(t)$ can exhibit
periodicity despite of $A<\theta$ due to stochastic
resonance (see Gammaitoni H\"anggi Jung and Marchesoni (1998)).

\begin{figure}[h]
\centering
\includegraphics[scale=0.4]{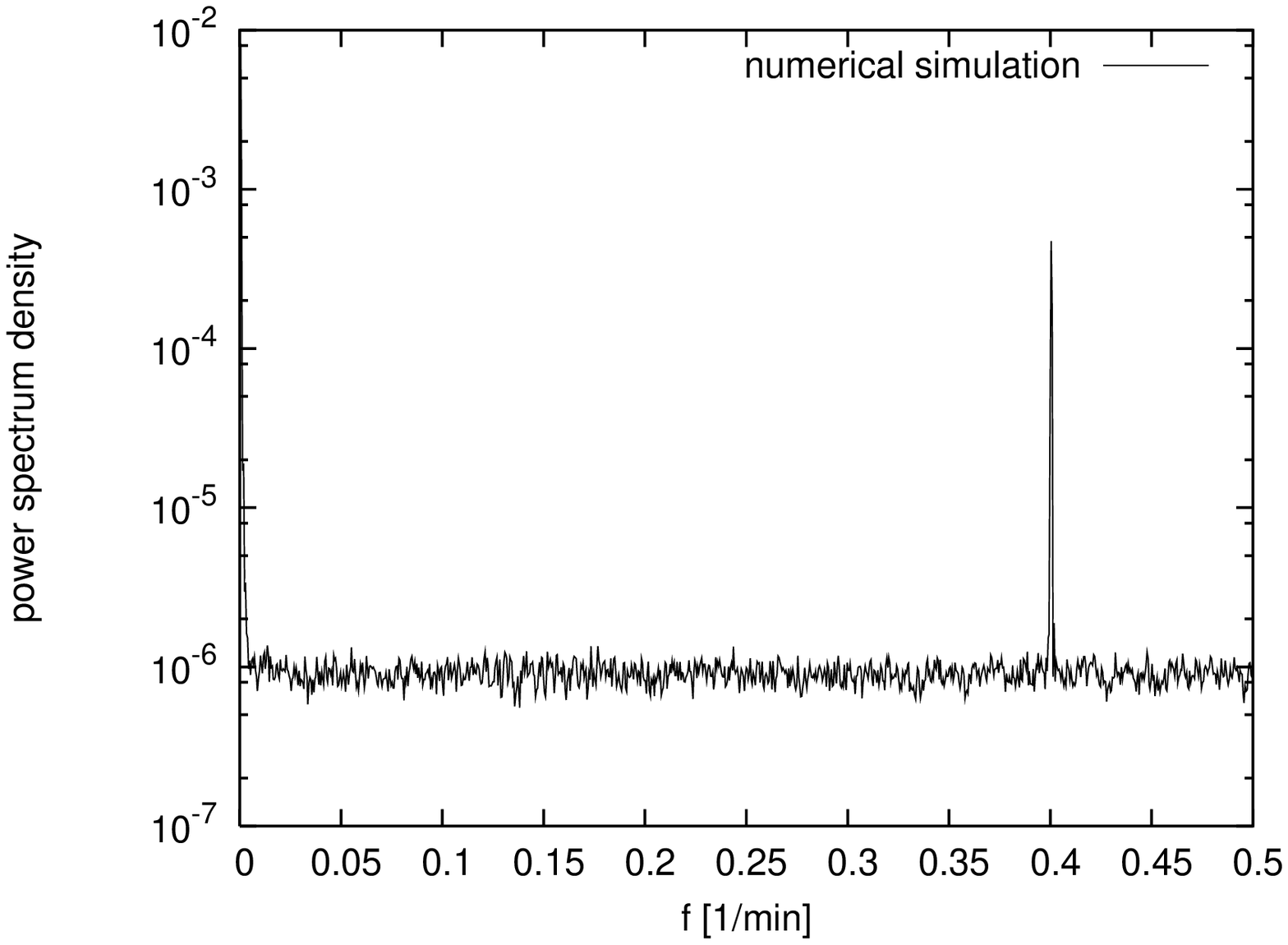}
\caption{Semi-log plots of the averaged power spectrum of $X(t)$ at
  $\sigma=0.3$, $\theta = 1.0$, $A=0.4$, and $f=0.2$. It has a peak at
  0.4.}
\label{fig:simulation}
\end{figure}

As shown in Fig. \ref{fig:simulation} it is found that the PSD has
some peaks from numerical simulations of the dealer model for
$\sigma>0$ and $A<\theta$. This peak is caused by stochastic resonance. 

\section{Discussion and Conclusion}
We empirically investigate the number of the tick quotes per unit time
in foreign currency market (USD/JPY). It is found that the power
spectrum densities of them for a period 1998 to 2000 all have a peak
at 0.4 [1/min]. From the results it is conclude that a periodical
action of dealers exists.

In order to explain this phenomena a simple dealer model based on the
double-threshold noisy devices is proposed. Under a hypothesis that
the mechanism of this periodicity is stochastic resonance the market
activity in the model shows periodicity due to uncertainty of dealers'
decision even though the exogenous periodical information is weaker
than the threshold for dealers to decide their action. In fact this
model is a feedforward one, however, real markets contains complicated
(positive and negative) feedbacks. The future work is to consider the 
feedbacks to improve the dealer model.

Moreover the source of this periodicity is open problem. One
possibility is an endogenous feedback mechanism of dealers as shown in
MT. The other is an exogenous periodical information as shown in this
paper. More detailed data analyses let us clarify the mechanism of
this phenomena. To consider this problem is expected to contribute to
a deep understanding of fluctuations and structure in the market.

\addcontentsline{toc}{section}{Index}
\flushbottom
\printindex

\end{document}